\documentclass[12pts]{article}
\usepackage{lineno}
\usepackage[dvips]{graphicx}
\usepackage[toc,page]{appendix}
\usepackage{graphicx}
\usepackage{amsmath}
\usepackage[varg]{txfonts}

\usepackage{longtable}

\usepackage[usenames]{color}
\usepackage{xspace}
\usepackage{siunitx}

\usepackage{natbib}
\begin{document}
\title{Whistler wave propagation through the ionosphere of Venus}

\author{
  F. J. P\'erez-Invern\'on$^{1}$,
 N. G. Lehtinen$^{2}$,
 F. J. Gordillo-V\'azquez$^{1}$,
  A. Luque$^{1}$. \\
\textit{$^{1}$Instituto de Astrof\'isica de Andaluc\'ia (IAA),} \\
   \textit{CSIC, PO Box 3004, 18080 Granada, Spain.} \\
\textit{$^{2}$Birkeland Center for Space Science, Institute of Physics and Technology,}  \\
\textit{University of Bergen, Norway.} \\
\footnote{Correspondence to: fjpi@iaa.es. 
Article published in Journal of Geophysical Research: Space Physics. doi.org/10.1002/2017JA024504 }
}
\date{}
\maketitle

\begin{abstract}

We investigate the attenuation of whistler waves generated by hypotetical venusian lightning occurring at 
the altitude of the cloud layer under different ionospheric conditions. We use the Stanford Full Wave Method (FWM) for stratified media of \cite{Lehtinen2008/JGR} to model wave 
propagation through the ionosphere of Venus. This method calculates the electromagnetic field created by an arbitrary source in a plane-stratified medium (i.e. uniform in the horizontal direction).
We see that the existence of holes in electronic densities and the magnetic field configuration caused by 
solar wind play an important role in the propagation of electromagnetic waves through the venusian ionosphere.

\end{abstract}

\section{Introduction}
\label{sect:intro}

Electrical discharges in planetary atmospheres produce electromagnetic waves that can be generally detected by spacecraft. 
In 1978, the Soviet Venera 11 and 12 detected bursts of very low frequency (VLF) pulses in their descent through the 
Venus atmosphere \citep{Ksanfomaliti1979/SAL, Ksanfomaliti1980/Natur}, 
similar to electromagnetic pulses generated by terrestrial lightning that can propagate up to some thousands of kilometers through
 the Earth-Ionosphere waveguide. Also in 1978 the Pioneer Venus Orbiter (PVO) detected VLF waves in several flybys over the 
Venus ionosphere, at altitudes from \SI{130}{km} to \SI{300}{km} \citep{Scarf1980/JGR, Strangeway1993/GRL,Strangeway1995/JATP, Strangeway2003/ASR}. These radio bursts, with a frequency  between \SI{85}{Hz} and \SI{115}{Hz}, were similar to whistler waves detected by 
spacecraft above electrical storms on Earth. \cite{Sonwalkar1991/JGR} studied the properties of these bursts, concluding that an important number of them could be caused by subionospheric lightning. However, other missions, such as Galileo and Cassini, did not reproduce these measurements, detecting only noise in 
the range of radio frequencies \citep{Gurnett1991/Sci, Gurnett2001/Natur}.
The Venus Express (VEX) probe, designed to study electromagnetic waves with a fluxgate magnetometer, was launched in 2005. This 
mission detected signals at altitudes around \SI{200}{km} in the range of frequencies between \SI{0}{Hz} and \SI{64}{Hz},
confirming the previous results of PVO \citep{Russell2013/ICA}. 

The lack of unambiguous optical signals from possible electrical lightning discharges on Venus keeps the origin of these PVO and VEX recorded whistler waves unclear. In 1975, the Venera 9 probe detected light from a storm system on Venus \citep{Krasnopolsky1980/CosmicRes}, whereas \cite{Hansell1995/Icarus} recorded optical signals using a ground-based telescope. However, no more optical results have been obtained since then \citep{Krasnopolsky2006/Icar, Garcia2013/GRL, Cardesin2016/ICA}. Moreover, some authors, as \cite{Cole1997/JGR}, claim that the plasma conditions in the ionosphere of Venus prevent the propagation of whistler waves generated by lightning. In addition, some studies proposed plasma instabilities (at much higher altitudes than the lightning, so that the propagation is not affected by the ionosphere) as  a possible source of the bursts observed by PVO, such as fluctuations driven by field-aligned beams \citep{Brace1991/SSR} or lower-hybrid-drift instabilities produced in zones where the plasma density is low and the background magnetic field is important \citep{Huba1992/JGR}. This hypothesis is consistent with the electron density fluctuations observed by PVO, as whistler waves do not perturb the plasma density \citep{Huba1992/JGR, Cole1997/JGR}.

Nevertheless, the existence of holes in the electronic and ionic density in the ionosphere of Venus together with a vertical magnetic field induced by solar winds \citep{Marubashi1985/JGR} can alter the plasma conditions facilitating the wave propagation through the ionosphere \citep{Ho1991/JGR}.
 
\cite{Huba1993/JGR} investigated electromagnetic wave propagation through the Venus ionosphere by solving the Maxwell equations in one dimension under several assumptions. They calculated the attenuation of whistler waves and concluded that the propagation of these waves is possible under ionospheric hole conditions. Since the method proposed by \cite{Huba1993/JGR}  was one-dimensional, it neglected the effect of refraction. In this paper, we apply a full-wave method for stratified media in order to study the propagation of waves at the frequencies recorded by PVO and VEX through the venusian atmosphere. We investigate the influence of wave refraction, atmospheric conditions, lightning characteristics and global lightning rate on the signals that can reach orbital altitudes and calculate the transfer function of the venusian ionosphere for each case. We obtain results consistent with \cite{Huba1993/JGR}.

\section{Plasma conditions in the ionosphere of Venus}
\label{sect:plasmaconditions}
Terrestrial lightning emits electromagnetic waves in the whole spectrum of radio frequencies \citep{Rakov2003/ligh.book}. The electromagnetic measurements taken by PVO and VEX spacecraft related to lightning events in Venus are inside the ELF (defined as 300--3000~Hz) and VLF (3~kHz--30~kHz) range of frequencies. On Earth, waves in this range of frequencies can travel through the ionosphere and magnetosphere as right-hand circularly polarized whistler-mode waves, following the geomagnetic field lines or irregularities in the electron and ion density profile.  During propagation, they may also undergo amplification  by interacting with energetic charged particles in the plasma. In the Earth ionosphere, attenuation in this range is usually low enough so that the waves are detectable from spacecraft. A wave is called ducted  when its propagation is restricted to  irregularities in the ionosphere with a layer or tube structure where the refractive index gradient allows this region to behave as a cavity that guides the radio signals. On the contrary, a wave is non-ducted when there is no such guiding irregularity, so that the wave suffers attenuation due to geometric divergence as the wave front expands. Ducted waves are not subject to such geometric attenuation because the wave front does not expand, being restricted to the transverse size of the irregularity. We should note here, however, that the geometric attenuation may also be small in the case of whistler waves in the Earth's ionosphere and magnetosphere, due to their propagation being restricted to the cone within the Storey angle. Some studies suggest that ducted wave are possible on the Earth magnetosphere \citep{Cerisier1974/JATP, Loi2015/GRL}. For an introduction to whistler-mode propagation through the ionosphere we refer to the textbooks \cite{Helliwell1965/book} and \cite{Budden1985/book}.

Whistler wave propagation through the ionosphere of Venus is determined by the background magnetic field and particle collision frequency profiles. The attenuation mostly depends on electron-neutral, ion-neutral and electron-ion collisions. Although electron-ion collision frequency is greater than electron-neutral collisions above 140 km under typical Venus ionospheric conditions, we have the opposite situation in ionospheric holes \citep{Huba1992/JGR, Huba1993/JGR}. For this reason, we have only included electron-neutral and ion-neutral collisions in our scheme. 
Electron-neutral and ion-neutral collision frequencies, denoted respectively as $\nu_{e}$ and $\nu_{i}$, depend on electron mobility $K_e$ and ion mobility $K_i$ as 

\begin{linenomath*}
\begin{equation}
\label{eq:P}
  \nu_{e,i} = \frac{e}{K_{e,i} m_{e,i}},
\end{equation}
\end{linenomath*}

where $e$ and $m_e$ are the electron charge and mass, respectively. We obtain the mobility of electrons in the ionospheric plasma from BOLSIG+ \citep{Hagelaar2005/PSST}. In the case of ions, we calculate the mobility of each considered ion following \cite{Mcdaniel1973/book}  and \cite{Borucki1982/Icarus} as

\begin{linenomath*}
\begin{equation}
\label{eq:P}
  K_{i} = 3.74 \times 10^{20} \left( M_i \xi \right)^{-0.5} n^{-1} \left[ cm^2 V^{-1} s^{-1} \right],
\end{equation}
\end{linenomath*}

where $n$ is the number density of the atmosphere at a given altitude, $\xi$ is the polarizability of CO$_2$, and $M_i$ is the reduced mass of the ion, related to the mass of each ion ($m_i$) and the mass of CO$_2$ molecules ($m_{CO_2}$) as

\begin{linenomath*}
\begin{equation}
\label{eq:P}
  M_i = \frac{m_i m_{CO_2}}{m_i + m_{CO_2}}.
\end{equation}
\end{linenomath*}

 \cite{Strangeway1996/JGR} obtained that the collisional Joule dissipation can be important in the lowe ionosphere of Venus for electric field amplitudes between 10~mV/m and 100~mV/m. However, fields obtained in our works are below these values. Therefore, we neglect the effect of the collisional Joule dissipation.

According to Pioneer Venus measurements taken above \SI{120}{km} altitude at nighttime conditions, O$_2^+$ is the main positive ion species between \SI{120}{km} and \SI{150}{km}, while O$^+$ is dominant at altitudes above \SI{150}{km} \citep{Taylor1979/Science, Bauer/ASR}. Electrons are the most important negatively charged particle species at these altitudes. The lack of measurements below \SI{120}{km} forces us to use models to estimate ion and electron atmospheric composition. According to \cite{Borucki1982/Icarus} and \cite{Marykutty2009/JGR}, negative and positive clusters are dominant charge carriers below \SI{60}{km} of altitude, while electrons and positive clusters are the main charged particles between \SI{60}{km} and \SI{75}{km}. We use the calculated electron and ion density profiles by \cite{PerezInvernon2016/JGR}, where electrons, O$_2^+$  and CO$_2^+$ ions are the dominant charged particles between the altitudes \SI{75}{km} and \SI{120}{km}. 

Both the orientation and value of the background magnetic field are crucial in the propagation of a wave through a plasma. Venus does not have a background magnetic field originating in its core, however, the solar wind interaction with its ionosphere induces a magnetic field of complex structure  \citep{Marubashi1985/JGR}. Whistler mode propagation is more feasible when the background magnetic field is parallel to the wave propagation direction. On the other hand, this propagation is prevented by a horizontal background magnetic field. According to PVO and VEX measurements, the observation of whistler waves coincided with vertical background magnetic field orientation and holes in the charged particle density profiles \citep{Taylor1979/Science, Strangeway2003/ASR, Russell2013/ICA}.

We can analyze the possible propagation modes using the Clemmow-Mullay-Allis (CMA) diagram for a two-component plasma \cite[p.~27]{Stix1992/Book}.  This diagram demonstrates different plasma wave modes at given frequency $\omega$, depending on  the relation between that frequency, the electron gyrofrequency  $\Omega_e$ and electron plasma frequency $\omega_{pe}$ determined by the plasma. Namely, the diagram shows different regions in which $\omega$ is smaller or greater than $\Omega_e$ and $\omega_{pe}$, explaining the propagation mode of the electromagnetic wave for each case.

For parameters relevant to this study the wave propagation is not very affected by ions, as in the case of a background magnetic field of 10 nT the lower hybrid frequency and the ion gyrofrequency are below 5 Hz and 2$\times$10$^2$ Hz, respectively.

The Budden parameters \citep{Budden1985/book} $Y=\Omega_e/\omega$ and $X=\omega_{pe}^2/\omega^2$ serve as the ordinate and the abscissa of the diagram, and the values of the parameters
\begin{linenomath*}
\begin{equation}
\label{eq:R}
R =1 - \frac{\omega_{pe}^2}{\omega\left(\omega-\Omega_e\right)} - \frac{\omega_{pi}^2}{\omega(\omega+\Omega_i)} ,
\end{equation}
\end{linenomath*}
\begin{linenomath*}
\begin{equation}
\label{eq:L}
L =1 - \frac{\omega_{pe}^2}{\omega\left(\omega+\Omega_e\right)} - \frac{\omega_{pi}^2}{\omega(\omega-\Omega_i)} ,
\end{equation}
\end{linenomath*} and
\begin{linenomath*}
\begin{equation}
\label{eq:P}
P =1 - \frac{\omega_{pe}^2 + \omega_{pi}^2 }{\omega^2} ,
\end{equation}
\end{linenomath*}

described in \cite{Stix1992/Book}, form the bounding surfaces that determine regions where the wave changes its propagation mode. An electromagnetic wave propagates in whistler mode when its frequency is lower than the plasma frequency ($\omega \ll \omega_{pe}$) and has a value between the ion gyrofrequency and the electron gyrofrequency  ($\Omega_i \ll \omega \ll \Omega_e$). These constrains are represented in the CMA diagram by regions 7 and 8  \cite[p.~27]{Stix1992/Book}, where the conditions

\begin{linenomath*}
\begin{equation}
\label{eq:P}
\frac{\omega_{pe}^2 + \omega_{pi}^2 }{\omega^2} > 1
\end{equation}
\end{linenomath*}

and

\begin{linenomath*}
\begin{equation}
\label{eq:P}
\Omega_{e} > \omega 
\end{equation}
\end{linenomath*}

are satisfied.

To analyze the case of whistler waves propagating through the venusian ionosphere we can assume an electron density peak of n$_{e_0}$ = 3$\times$10${^4}$~cm$^{-3}$ at nighttime conditions \citep{Bauer/ASR} and O$^+$ ions as the main positive charged particle. We can calculate the electron plasma frequency $\omega_{pe}$, ion plasma frequency $\omega_{pi}$, electron gyrofrequency $\Omega_{e}$ and ion gyrofrequency $\Omega_{i}$ for a wave of frequency \SI{100}{Hz} with different background magnetic fields. We estimate that whistler-mode propagation conditions in regions 7 and 8 of the CMA diagram are satisfied by the wave of \SI{100}{Hz} when the vertical component of the background magnetic field is greater than 3.57~nT if the electron density peaks at 3$\times$10${^4}$  cm$^{-3}$. We also estimate that  $\omega_{pe}>\SI{100}{Hz}$ even in electron density holes with the electron density peak  reduced up to $\sim$8 orders of magnitude (i.e., a factor of $\sim$$10^8$), establishing this density value  as the lowest at which non-ducted whistler-mode propagation through the venusian atmosphere is possible. The upper electron density values for a given background magnetic field, on the other hand, are constrained by the condition that attenuation due to electron collisions is low enough for the signals to be detectable, which we are going to demonstrate.

\section{Model}
\label{sect:model}

\begin{figure}
\includegraphics[width=0.9\columnwidth]{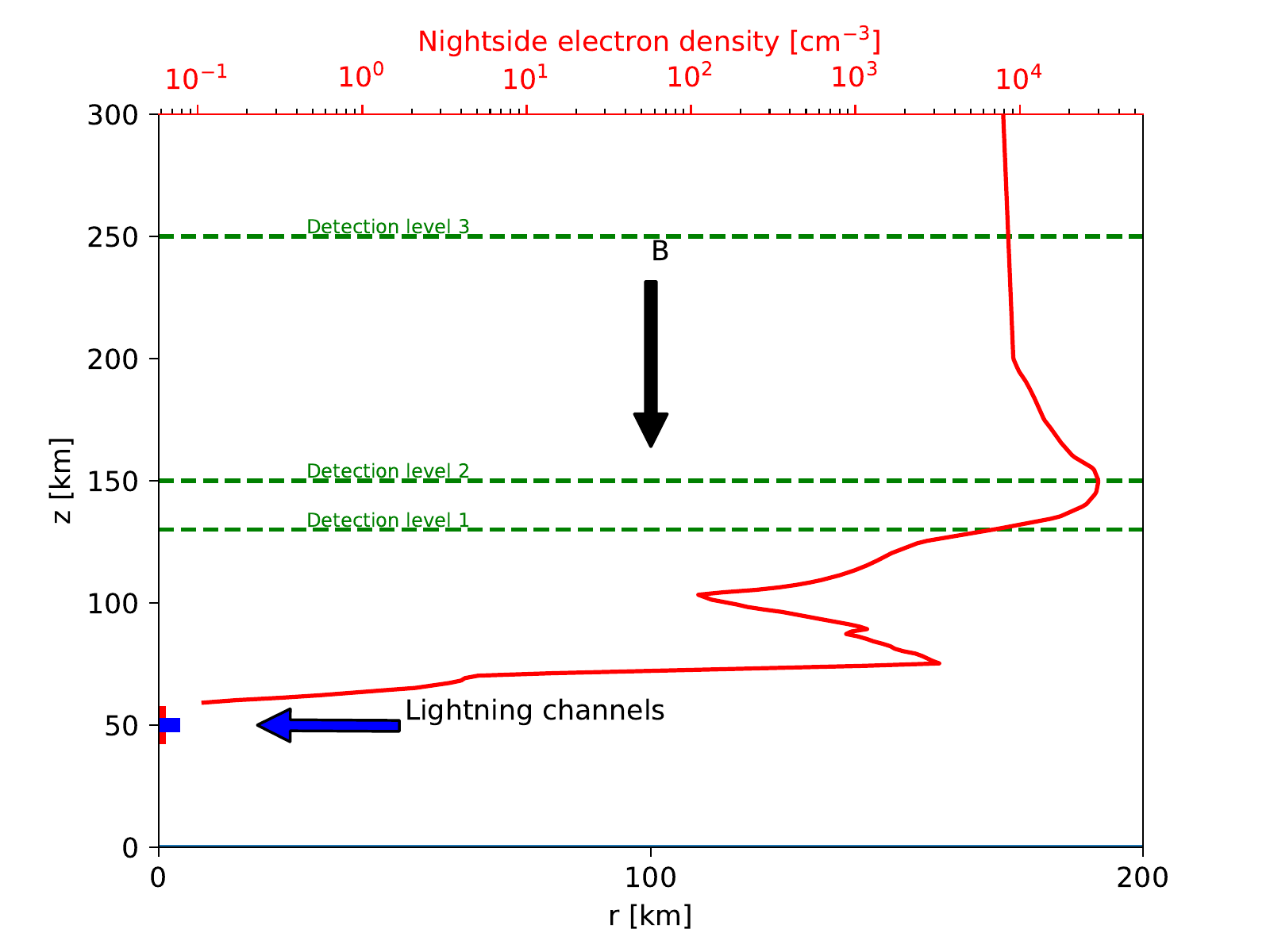}
\caption{\label{fig:scheme}
Geometrical scheme in cylindrical coordinates and ambient concentration of electrons in the nightside of Venus (red line) \citep{Borucki1982/Icarus, Bauer/ASR,PerezInvernon2016/JGR}. Lightning channels can be vertical (red) or horizontal (blue). We calculate lightning-produced electromagnetic fields at detection levels (green dashed lines) to compare with observations. The background magnetic field is vertical to ground everywhere.
}
\end{figure}

\subsection{Stanford Full Wave Method applied to Venus}
The Stanford Full Wave Method (StanfordFWM) described in \cite{Lehtinen2008/JGR}, \cite{Lehtinen2009/GRL} and \cite{Lehtinen2010/JGRA} solves the Maxwell equations and calculates all the electromagnetic field components in the frequency domain for a wave generated in and propagating through a stratified medium. We apply this model to calculate the propagation of a wave created by lightning through the ionosphere of Venus during nighttime conditions. We study the propagation of waves with frequencies measured by PVO and VEX, between \SI{5}{Hz} and \SI{100}{Hz}, for different atmospheric conditions. Since the irregularities that produce ducted waves cannot be defined in this stratified medium scheme, we restrict our study to non-ducted waves. The dielectric permittivity tensor is assumed to be uniform within each layer. Its value depends on the charged particle densities and collision frequencies detailed in section \ref{sect:plasmaconditions}, as well as on the background magnetic field \citep{Lehtinen2008/JGR, Lehtinen2009/GRL, Lehtinen2010/JGRA}.

As we mentioned in section~\ref{sect:plasmaconditions}, the existence of vertical background magnetic fields together with holes in the charged particle density profiles allows whistler-mode propagation. In this model, we calculate wave propagation from a source through the venusian ionosphere for vertical background magnetic fields with magnitudes between 0~nT and 40~nT. Regarding the charged particle density holes, the reduction of the density can be greater than 3~orders of magnitude \citep{Marubashi1985/JGR}. We calculate wave propagation through holes where the charged particle density reduction is between 0 and 8 orders of magnitude. This lowest electron density value is probably far below what is expected, but we include it in our calculations to emphasize how low the electron density must be to explain the detected values.

We consider vertical and horizontal intra-cloud (IC) discharges with a length of 7 km and located at 45 km of altitude, with a waveform specified by a bi-exponential function \citep{PerezInvernon2016/JGR}:

\begin{linenomath*}
\begin{equation}
\label{eq:current}
I(t) = I_0 \left(\exp(-t/\tau_1) - \exp(-t/\tau_2)\right),
\end{equation}
\end{linenomath*}
where $\tau_1$ = 1 ms and  $\tau_2$ = 0.1 ms are, respectively, the total and the rise times of the electric current $I$. The parameter $I_0 = Q / (\tau_1 - \tau_2)$, where Q is the total transferred charge, is obtained following the approximation of \cite{Maggio2009/JGR} for an extreme case where the total electromagnetic energy released by the discharge is 2 $\times$ 10$^{10}$~J. The results can be scaled to cases with total released energy between 8 $\times$ 10$^{8}$~J and 10$^{10}$~J \citep{Krasnopolsky1980/CosmicRes}, or to Earth-like intra-cloud lightning energies between 10$^{6}$~J and 10$^{7}$~J \citep{Maggio2009/JGR}.

The FWM can be applied to calculate wave propagation in different coordinate systems depending on the background magnetic field inclination. Since we model wave propagation in the presence of a vertical magnetic field, we use a cylindrically symmetrical scheme. Finally, results are transformed to a cartesian coordinate system as follows. We neglect the planetary curvature, as the considered wave frequencies are above the Schumann resonances, produced at frequencies between 9~Hz and 24~Hz  \citep{Simoes2008/JGR}.

In the case of the horizontal discharge, the source has two spatial components, denoted as $I_x$ and $I_y$. Electromagnetic wave propagation from an horizontal discharge channel can be modeled with FWM taking advantage of the source expansion in axial harmonics, provided that the dielectric permittivity is axisymmetric. We use the fact that the Fourier transform of any function defined in the space coordiantes $x$ and $y$ that depends on the radial distance $r$ of form

\begin{linenomath*}
\begin{equation}
\label{eq:function}
A(x,y)=A(r) \exp\left( i \thinspace n\thinspace \phi \right),
\end{equation}
\end{linenomath*}

where $i$ is  the imaginary unit and $n$ is an integer, is

\begin{linenomath*}
\begin{equation}
\label{eq:transform}
A(k_x,k_y)=\frac{2 \pi}{i^n} A_n(k) \exp\left(i\thinspace n\thinspace \chi\right),
\end{equation}
\end{linenomath*}

where the 2D Fourier transform is defined as

\begin{linenomath*}
\label{eq:fourier}
\begin{equation}
A(k_x,k_y)= \int \exp\left(-i\thinspace(k_xx+k_yy)\right)A(x,y) \thinspace dx \thinspace dy ,
\end{equation}
\begin{equation}
 A(x,y)  = \int \exp\left(i\thinspace(k_xx+k_yy)\right) A(k_x,k_y) \frac{dk_x \thinspace dk_y}{(2 \pi)^2} , 
\end{equation}
\end{linenomath*}

with the integrals having infinite limits. The polar coordinates used above are

\begin{linenomath*}
\label{eq:coord}
\begin{equation}
x=r \cos(\phi), y=r \sin(\phi)
\end{equation}
\begin{equation}
k_x=k\cos(\chi), k_y=k\sin(\chi),
\end{equation}
\end{linenomath*}

while $ A_n(k)$ is the Hankel transform of $A(r)$, defined as

\begin{linenomath*}
\label{eq:hankel}
\begin{equation}
A_n(k) = \int_0^\infty A(r) J_n(k r) r \thinspace dr
\end{equation}
\begin{equation}
 A(r)   = \int_0^\infty A_n(k) J_n(k r) k \thinspace dk ,
\end{equation}
\end{linenomath*}

where $J_n(k r)$ is the nth order Bessel function of the first kind.

The function $A$ is a scalar, which can be any of the components of a vector, e.g., $I_x$, $I_y$ or $I_z$. However, the FWM takes as an  input of the source the values of $I_x (k)$, $I_y (k)$ and $I_z (k)$ at $\chi=0$ \citep{Lehtinen2008/JGR, Lehtinen2009/GRL, Lehtinen2010/JGRA}. To calculate the fields we need to know the angular dependence $I_x(\chi)$, $I_y(\chi)$ and $I_z(\chi)$. The input to the FWM requires that the polar components of the current in $(k_x,k_y,z)$-space satisfy $(I_k,I_{\chi},I_z)  \sim \exp(i m \chi) $. Then the polar components of the electromagnetic fields in $(k_x,k_y,z)$-space, also have the same $\chi$-dependence (due to linearity and axial symmetry of the dielectric permittivity). These can be then converted to fields in $(x,y,z)$-space at $\phi$=0.

 The geomtetrical scheme, shown in figure~\ref{fig:scheme}, can be summarized as follows:

\begin{enumerate}  
\item Cylindrical coordinate system, where the $r$ axis is parallel to the ground and the axis of symmetry, denoted as $z$, points towards nadir. We do not take into account the planetary curvature, as our purpose is to calculate upward propagation from a stroke locally.  Results are transformed to cartesian coordinates before plotting.

\item Background magnetic field vertical to the ground everywhere.

\item Two different lightning-channel inclinations,  parallel and vertical to the ground. The source is located in the $z$ axis. The direction and propagation of the emitted electromagnetic wave will be obtained for each case.

\end{enumerate}

\subsection{Method of comparison with measurements}
The Full Wave Method allows us to obtain the electromagnetic wave components at a given altitude and horizontal distance from the source. However, spacecraft as Pioneer Venus Orbiter and Venus Express did not measure electromagnetic field components directly, but the power spectral density within a frequency band during a certain time interval \citep{Scarf1980/JGR, Strangeway2003/ASR, Russell2013/ICA}. In this section, we will explain an approximate method to compare our results with the available measurements at frequencies between 5 Hz and 100 Hz.

The $\omega$-space current moment is the Fourier image of equation (\ref{eq:current}):

\begin{linenomath*}
\begin{equation}
\label{eq:currentw}
K( \omega ) = L I_0 \left(\frac{1}{\frac{1}{\tau_1} - i\omega} - \frac{1}{\frac{1}{\tau_2} - i\omega} \right),
\end{equation}
\end{linenomath*}

where $L$ is the dipole length. 

We use the FWM to calculate the cartesian electromagnetic field components \\ $\mathbf{F} (\omega) = (E_x, E_y, E_z, H_x, H_y, H_z)$ at a given altitude for a particular frequency. The transfer functions can be defined as the ratio of absolute values between each field component and the source:

\begin{linenomath*}
\begin{equation}
\label{eq:T}
T_j( \omega ) =  \frac{\left|F_j(w)\right|}{\left|K(w)\right|},
\end{equation}
\end{linenomath*}

where $j$ denotes the electromagnetic field component.

To relate the experimentally observed power spectrum of electromagnetic radiation to the energy spectrum of the source given by $K(\omega)$, we need to make some assumptions regarding the lightning temporal distribution. The simplest approximation is to assume a constant rate $\nu$ of planetary lightning strokes per second with similar characteristics, neglecting stationallity and lightning occurrence correlation. Then the power spectral density for each electromagnetic field component is

\begin{linenomath*}
\begin{equation}
\label{eq:PSD}
S_j( \omega ) =  \nu \left|T_j(w)\right|^{2} \left|K(w)\right|^{2}.
\end{equation}
\end{linenomath*}

\cite{Russel1989/GRL} obtained an approximate value for the rate $\nu$ of 80 flashes per second with a total released energy of 2 $\times$ 10$^{10}$~J per flash. We will normalize our computations to a value of $\nu$=1 flash per second. By scaling our results we can then obtain the rate that best agrees with the measured power spectral density.

This way to calculate the power spectral density assumes an approximate lightning flash rate value that is constant in time. However, both PVO and VEX spacecraft recorded time variations of the power spectral density that can be due to several factors, as for example, a temporal dependence of the lightning distribution or due to some instrumental aspects.

The Pioneer Venus Orbiter was equipped with the Orbiter Electric Field Detector (OEFD), a plasma wave instrument with four band-pass channels centered at 100~Hz, 730~Hz, 5.4~kHz, and 30~kHz \citep{Scarf1980/JGR,Colin1980/JGR}. This instrument had an antenna \citep{Russell1990/ASR} that recorded electric field components on a plane, hence the changes of the angle between the direction of the wave propagation and the plane of measurement during the time of observation could entail a temporal variation in the measured power spectral density. In this work, we will compare our calculated power spectral density with PVO peak measurements at altitudes of 130~km, 150~km \citep{Strangeway2003/ASR} and 250~km \citep{Scarf1980/JGR} in the band-pass channel centered at 100~Hz, where whistler waves were observed.

The Venus Express instrumentation was different to that onboard PVO, as it recorded magnetic field components instead of electric field components. The VEX spacecraft was equipped with a magnetometer system (MAG) consisting of two sensors for magnetic field magnitude and direction  \citep{Titov2006/CR}. We will compare our power spectral density calculations with the measurements taken by VEX during nightside observations at 250~km altitude \citep{Russell2013/ICA}. Interestingly, VEX recorded transverse right-handed guided waves consistent with whistler mode propagation.

\section{Results}
\label{sect:results}

\begin{figure}
\includegraphics[width=0.9\columnwidth]{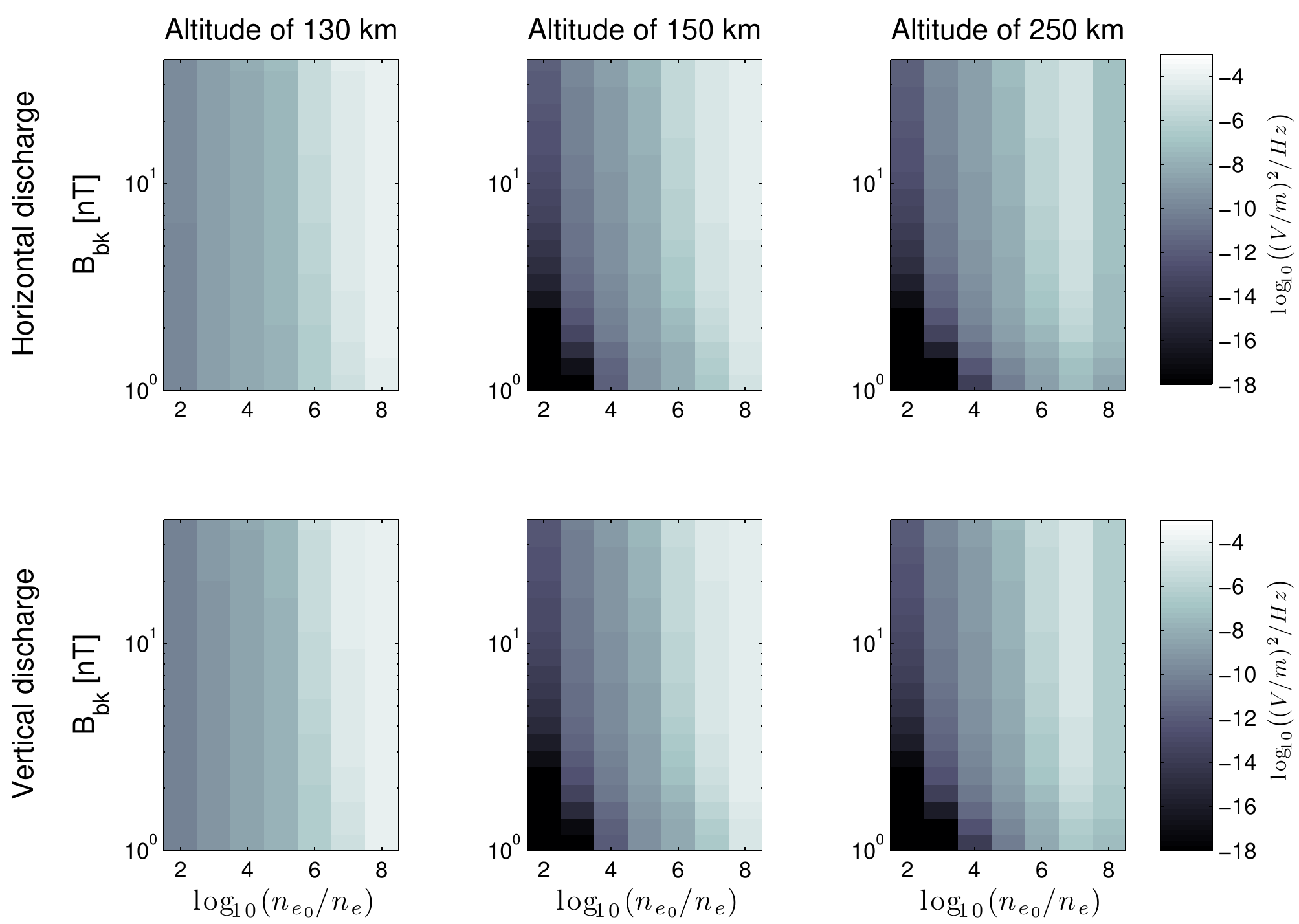}
\caption{\label{fig:PVO}
Maximum power spectral density (PSD) in (V/m)$^2$Hz$^{-1}$ calculated from the $E_x$ and $E_y$ components at different altitudes and resulting from a lightning rate of 1 stroke per second for horizontal discharges (first row) and vertical discharges (second row). The axes are the background magnetic field and the ionospheric hole magnitude, defined as the reduction of the electron density compared to the background value which peaks at around 3 $\times$ 10$^4$~cm$^{-3}$. This definition of the x-axis alludes explicitly to the holes detected by PVO. We plot results for background magnetic fields greater than 1 nT.
}
\end{figure}

\begin{figure}
\includegraphics[width=0.9\columnwidth]{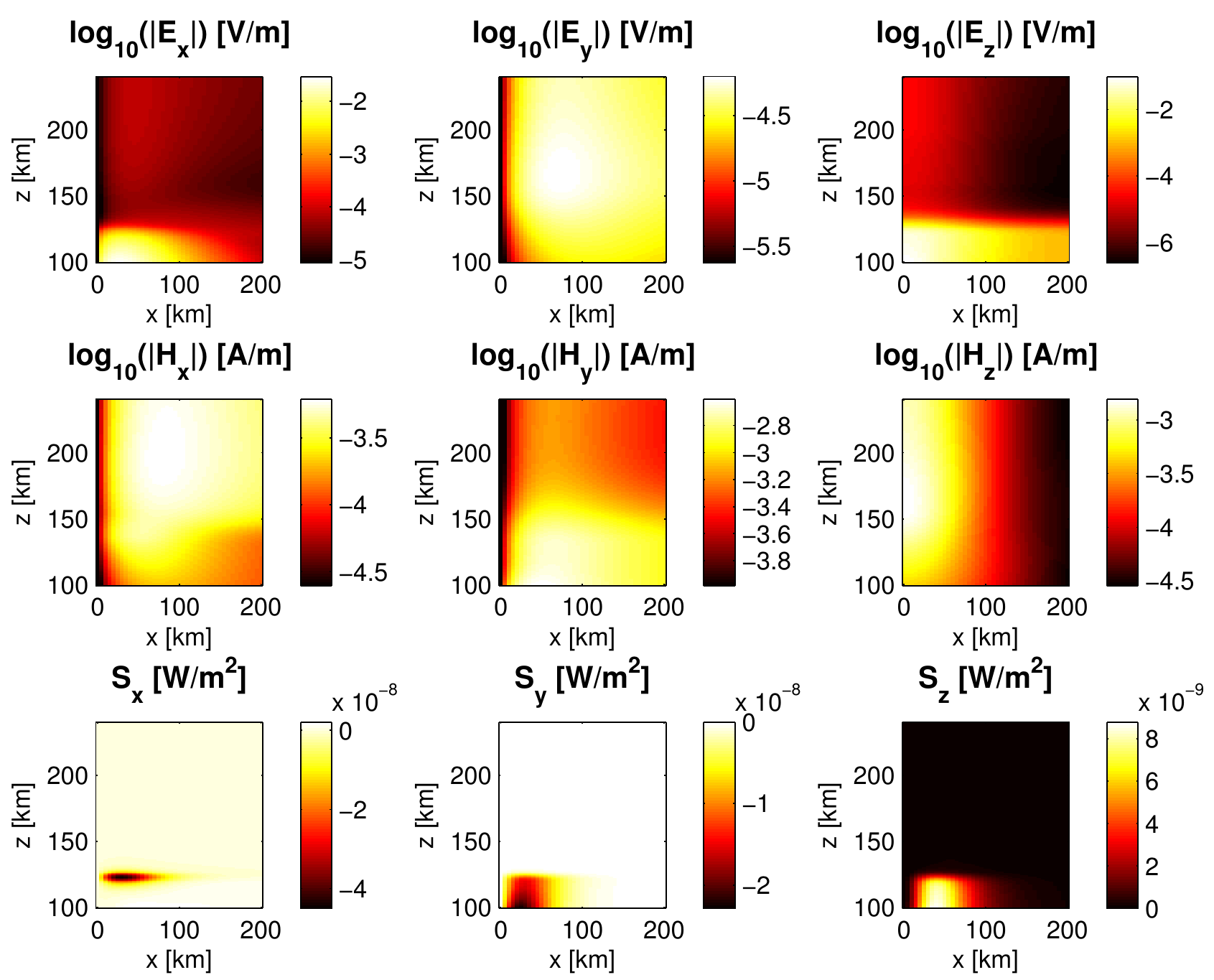}
\caption{\label{fig:PVO_vertical}
 Time averaged electromagnetic field and Poynting vector components produced by a vertical discharge in the ionosphere of Venus under a background magnetic field of 20 nT and a reduction of 5 orders of magnitude in the electron and ion densities. 
The vertical lightning discharge is located at 45 km of altitude and produces a wave with a frequency of 100 Hz. This plot shows atmospheric regions above 100 km of altitude.
}
\end{figure}

\begin{figure}
\includegraphics[width=0.9\columnwidth]{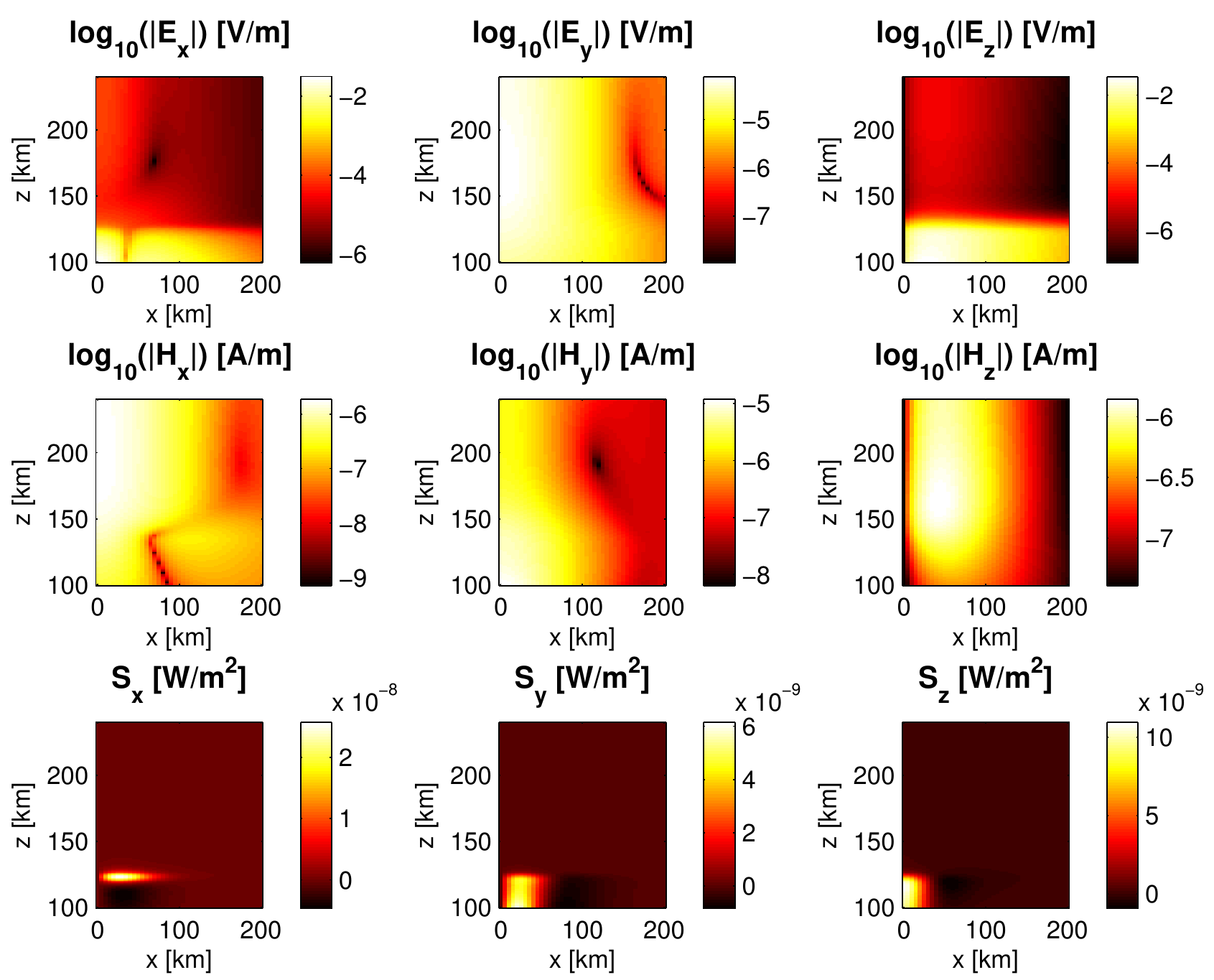}
\caption{\label{fig:PVO_horizontal}
 Time averaged electromagnetic field and Poynting vector components produced by a horizontal discharge in the ionosphere of Venus under a background magnetic field of 20 nT and a reduction of 5 orders of magnitude in the electron and ion densities. In this case, 
The horizontal lightning discharge is a dipole located at 45 km of altitude and contained in the plane x-z. The source produces a wave with a frequency of 100 Hz. As in figure \ref{fig:PVO_vertical}, this plot shows atmospheric regions above 100 km of altitude.
}
\end{figure}

\begin{figure}
\includegraphics[width=1\columnwidth]{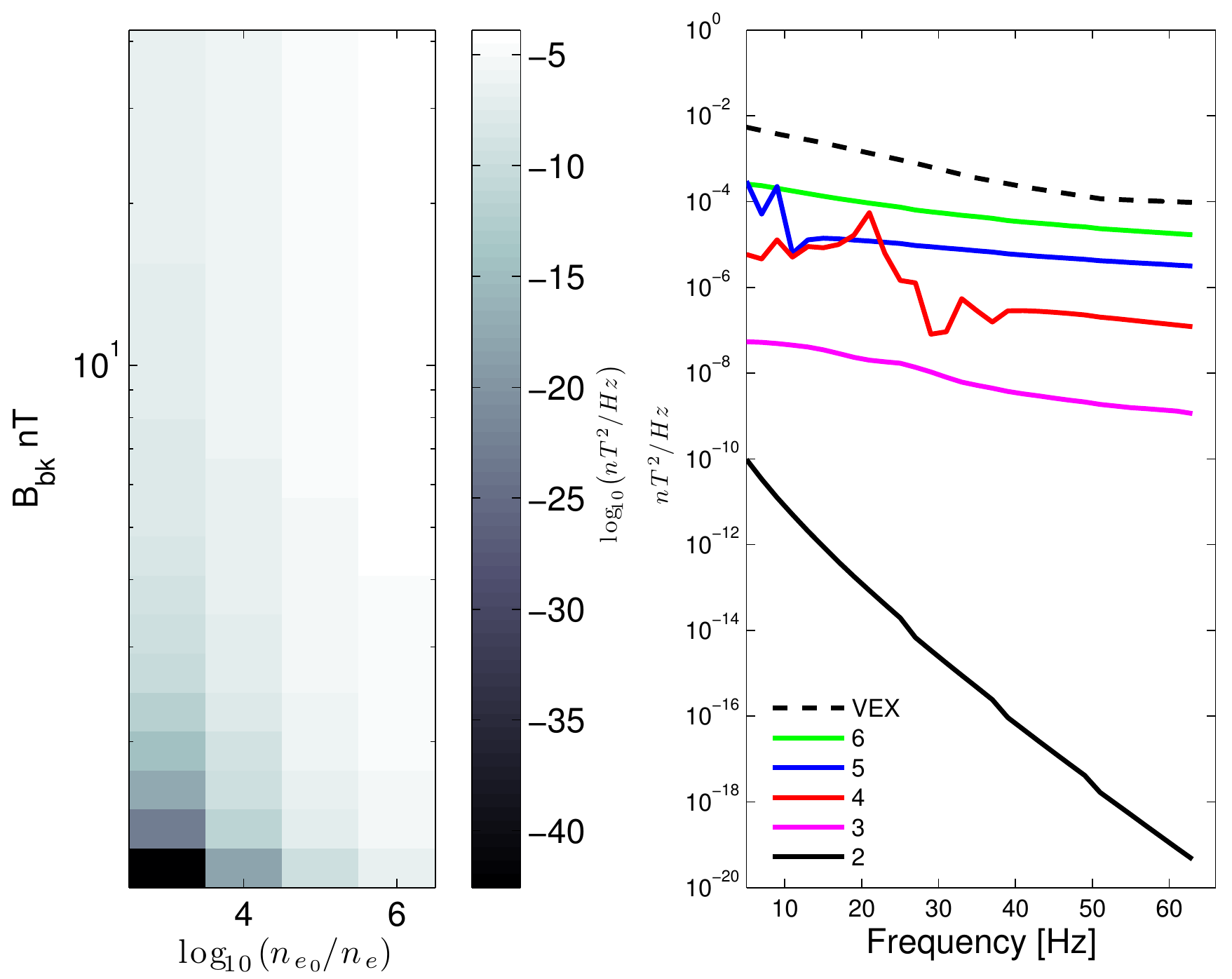}
\caption{\label{fig:VEX}
(left) Power spectral density at 250 km of altitude in (nT)$^2$Hz$^{-1}$ calculated at 40~Hz from the $B_x$ and $B_y$ components at different altitudes and resulting from a lightning rate of 1 stroke per second for vertical discharges. The axes are the same as in figure~\ref{fig:PVO}. We also plot results for background magnetic fields greater than 1 nT. (right) Calculated power spectral density for different frequencies and a fixed intermediate background magnetic field of 15~nT, where the number in the inset indicate the order magnitude of the reduction in the electron and ion density peak, n$_{e_0}$. The spectrum recorded in the nightside of Venus by VEX is also shown (dashed line) \citep{Russell2013/ICA}. Results below 24~Hz could change with the inclusion of the planetary curvature as a consequence of the Schumann resonances.
}
\end{figure}

\subsection{Electric field}

We will first analyze the power spectral density of electric field component at 100 Hz. Figure \ref{fig:PVO} shows such results at different altitudes for horizontal and vertical lightning discharges with a rate of 1 stroke per second and a total released energy per stroke of 2 $\times$ 10$^{10}$ J. These calculations are performed for different background magnetic field (vertical axis) and different reductions in the peak electron and ion density (horizontal axis) in the charged particle density profiles (see the figure caption for the definition). 

The maximum power spectral densities measured by PVO at 100~Hz between 130~km and 250 km of altitudes were around 10$^{-4}$ (V/m)$^2$Hz$^{-1}$ \citep{Scarf1980/JGR,Strangeway2003/ASR}. The second row of figure \ref{fig:PVO} shows that a lightning rate of 100 flashes per second would produce a power spectral density of around 10$^{-4}$ (V/m)$^2$Hz$^{-1}$ in the presence of a background magnetic field between 5~nT and 40~nT and a reduction in the electron and ion profile of ~10$^5$ and~10$^6$. The expected power spectral density decreases steeply when the hole is less pronounced. It is interesting to note that the calculated power spectral density at 250 km of altitude also decreases in the case of a depletion of 10$^8$ in electron density. This effect can be due to the lack of electrons creating unfavorable conditions to whistler mode propagation.

Figures \ref{fig:PVO_vertical} shows the electromagnetic field and Poynting vector components produced by a vertical lightning discharge in the ionosphere of Venus. According to the spatial distribution of the horizontal components of the electromagnetic field ($E_x$ and $E_y$), whistler mode propagation occurs at horizontal distance between 10 km and 100 km from the source. The Poynting vector components indicate the direction of the energy flux and the region of maximum absorption, at altitudes around 125 km. This altitude coincides with the largest enhancement in the electron density, where the induced current in the ionosphere shields the electromagnetic field and influences the flux direction. 

The electromagnetic field and Poynting vector components produced by a horizontal lightning discharge are shown in figure \ref{fig:PVO_horizontal}. The region of maximum absorption is also located at around 125 km of altitude. In this case, the whistler wave traverses the ionosphere traveling along a vertical column right above the source. The vertical component of the Poynting vector exhibits negative values at an horizontal distance of $\sim$50 km from the lightning discharge, indicating a downwards flux of energy produced by the reflection of the wave in the lower ionosphere.

\subsection{Magnetic field}

In this section we calculate the power spectral density from the calculated magnetic field component calculations in the range of frequencies between 5~Hz and 64~Hz, always above the lower hybrid frequency. Figure \ref{fig:VEX}a shows the maximum calculated power spectral density at 40~Hz and 250~km altitude for vertical lightning discharges assuming a rate of 1 stroke per second and a total released energy per stroke of 2 $\times$ 10$^{10}$~J. Again, the result is calculated for different background magnetic field values (vertical axis) and different reductions of the electron and ion densities (horizontal axis) in the ionosphere of the planet. 

In the case of the VEX spacecraft, power spectral measurements were generated using a magnetometer in the range of frequencies between 0~Hz and 64~Hz. We compare our calculations with measurements of transverse right-handed guided waves recorded by VEX during its nightside observation on June~9, 2006 \citep{Russell2013/ICA}. According to the data analysis performed by \cite{Russell2013/ICA}, the maximum power spectral densities at 40~Hz was around 10$^{-2}$ (nT)$^2$Hz$^{-1}$. However, the calculated power spectral density in figure \ref{fig:VEX}a suggests that a rate of around 100 flashes per second and a reduction of 5 orders of magnitude in the electron and ion density are necessary to produce the VEX observed power spectral density. As in the previous case, an increase in the charged particles density would produce more wave attenuation.

Figure \ref{fig:VEX}b shows the frequency dependence of the calculated spectral density for different hole magnitudes and a background magnetic field of 15~nT, together with the VEX recorded spectrum in the nightside by VEX \citep{Russell2013/ICA}. We see that different electron and ion densities cause different frequency-dependent attenuation in the propagating wave. Venus Express recorded radio signals for different atmospheric conditions \citep{Russell2013/ICA}, obtaining different spectra of the transverse right-handed guided wave during daytime and nighttime conditions. The downward slope exhibited by the spectra for hole magnitudes of 2, 4 and 5 are similar to the measurements taken by VEX in the venusian nightside \citep{Russell2013/ICA}. If we compare the curves in figure \ref{fig:VEX}b for different electron densities (hole magnitude) and curves obtained by VEX for different conditions, we see that the spectrum changes depending on the profile or the presence of a hole. Venus Express was not equipped with an instrument to measure plasma densities. However, our results and available VEX measurements suggest that there exists a direct relation between the spectrum of the wave and the profile of charged particles, with a flatter spectrum for higher electron densities.

\subsection{Transfer function}
\begin{figure}
\includegraphics[width=1\columnwidth]{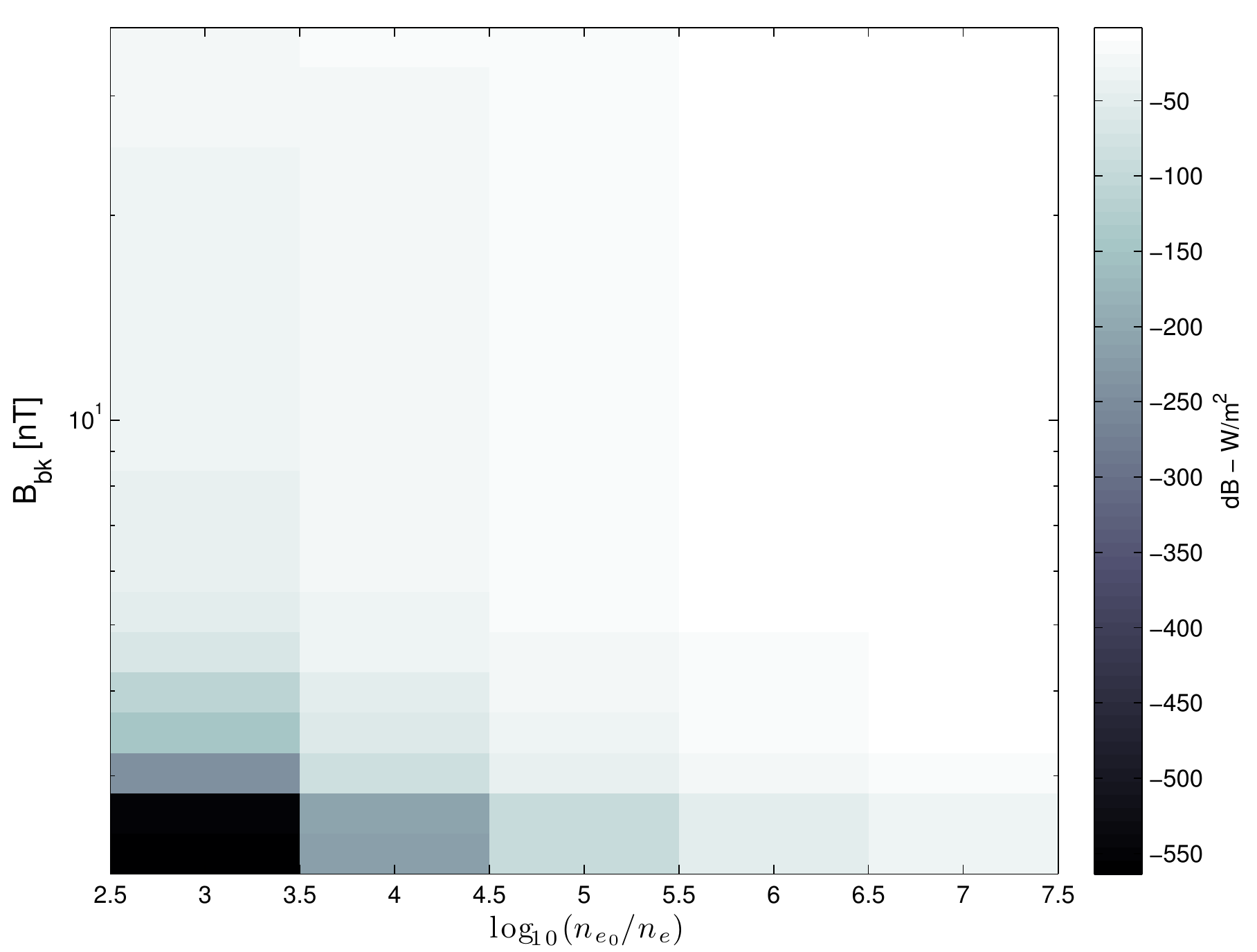}
\caption{\label{fig:transfer_function}
Attenuation of the electromagnetic signal (in dB) at 100~Hz after traversing the venusian ionosphere from 75 km to 250~km altitude under a background magnetic field greater than 3 nT. The axes are the same as in Figure~\ref{fig:PVO}. Here we plot values of the transfer function for different orders of magnitude in the reduction of the electron and ion density peak.
}
\end{figure}
We can calculate the attenuation in decibels for different atmospheric conditions (background magnetic field and electron density) using the time-averaged Poynting vector $\mathbf{S} = \frac{1}{2} \Re (\mathbf{E} \times \mathbf{H}^{*})  $, where $ \Re (x)$ stands for the real part of $x$. The attenuation is $A = 10 \log_{10} \left(\frac{S}{S_{0}}\right)$, where $S$ is the module of the Poynting vector at 250 km of altitude, and $S_0$ is the module of the Poynting vector at a reference level (calculated at 75 km of altitude).

Figure \ref{fig:transfer_function} shows the calculated $A$ for a 100~Hz electromagnetic wave traversing the venusian atmosphere altitudes from 75~km to 250~km for holes magnitudes between 3 and 7. In the vertical axis, we plot background magnetic fields greater than 3 nT. In the case of a background magnetic field lower than this value or hole magnitudes smaller than 3 the transfer function becomes almost zero. 

It can be seen that an ionosphere without holes and background magnetic field produces the strongest attenuation. However, the existence of holes together with background magnetic field enhance wave propagation.

\subsection{Ducted vs unducted propagation}
\label{sec:ducted}

Let us estimate the effects of the geometric attenuation in unducted case compared to ducted propagation. The unducted wave energy at the spacecraft altitude goes through the area which is of the order $D^2$ where $D\sim200$~km is the vertical distance between the source and the observing spacecraft. This area may be less if we take into account that the whistler wave group velocities are contained within the Storey cone around the background magnetic field \citep{Helliwell1965/book}. The geometric attenuation factor is therefore at most $D^2/a_\mathrm{duct}$, where $a_\mathrm{duct}$ is the cross-sectional area of the duct. Thus, for realistically sized ducts of $>$10~km transverse size, the ducted results would be about $10^2$--$10^3$~higher than those in the unducted case.

\section{Conclusions}
\label{sect:conclusions}

Very Low Frequency (VLF) waves recorded by Pioneer Venus Orbiter (PVO) and Venus Express (VEX) suggest the existence of lightning in Venus \citep{Taylor1979/Science, Scarf1980/JGR, Strangeway2003/ASR, Russell2013/ICA}. However, the source of these VLF signals is not completely clear since the altitude where they are produced is not known. In this work, whistler-wave propagation through the ionosphere of Venus has been studied using a Full Wave Method for stratified media \citep{Lehtinen2010/JGRA}. We have assumed that lightning in the venusian cloud layer act as the emitting source of VLF. The wave propagation through the atmosphere of Venus is investigated under certain conditions, like the presence of a background magnetic field \citep{Marubashi1985/JGR} and a reduction in the electron and ion densities \citep{Ho1991/JGR}.

The Full Wave Method calculates the electromagnetic field at a given altitude for given source characteristics and atmospheric conditions such as background magnetic field and electron and ion densities, thus taking care of possible absorption, refraction and reflection of the wave \citep{Lehtinen2008/JGR, Lehtinen2009/GRL, Lehtinen2010/JGRA}. We have made realistic assumptions about the lightning characteristics \citep{Krasnopolsky1980/CosmicRes, PerezInvernon2016/JGR} and the atmospheric conditions \citep{Borucki1982/Icarus, Bauer/ASR, Marubashi1985/JGR, Ho1991/JGR, Marykutty2009/JGR, PerezInvernon2016/JGR}, on the basis of which we obtained both the attenuation of whistler waves and the expected time-averaged power spectral density as a function of a global lightning rate. According to our results, unducted whistler-wave propagation is possible under the existence of local and temporal reductions in the electron and ion density and an induced background magnetic field greater than 4~nT and perpendicular to the wave propagation. These results are consistent with previous studies \citep{Huba1993/JGR}.

In addition, we have compared our estimates of the power spectral density with available measurements recorded by PVO and VEX. We have estimated the global flash rate (lightning flashes per second) needed to reproduce the observations in the presence of ionospheric holes. For lightning with a total energy released of 2 $\times$ 10$^{10}$ J \citep{Krasnopolsky1980/CosmicRes}, the required number of lightning per second is of the order of 100, as in \citep{Russel1989/GRL}. 
However, for other (more realistic) energies of the order of average terrestrial lightning with total energy released of 10$^{7}$ J, the needed number of lightning per second must be as high as around 10$^{6}$. This unrealistic value leads us to suggest that, if the observed signals are generated by lightning, either the average energy released by venusian lightning is considerably greater than the terrestrial lightning energies, or ducted wave propagation is common in the ionosphere of Venus, allowing VLF waves to propagate without suffering  geometric attenuation, which would increase the observed power by an estimated factor between $10^2$ and $10^3$.

Finally, we have evaluated the relation between the electron density in the ionosphere and the dispersion of the propagating wave and found values that are consistent with VEX results. 

Future missions with dedicated instrumentation to take precise measurements of plasma parameters together with radio wave signals could be useful to determinate the source of  the radio signals observed by the PVO and VEX missions. Furthermore, measurements of the wave attenuation dependence with frequency could provide useful information about the venusian atmosphere composition.

\section*{Acknowledgement}
This work was supported by the Spanish Ministry of Science and Innovation, MINECO under projects ESP2015-69909-C5-2-R and FIS2014-61774-EXP, by the Research Council of Norway under contracts 208028/F50, 216872/F50 and 223252/F50 (CoE),  and by the EU through the FEDER program. FJPI acknowledges a PhD research contract, code BES-2014-069567.  NL was supported by the European Research Council under the European Union’s Seventh Framework Programme (FP7/2007-2013)/ERC grant agreement n.~320839.  AL was supported by the European Research Council (ERC) under the European Union’s H2020 programme/ERC grant agreement n.~681257. All data used in this paper are directly available after a request is made to authors F.J.P.I (fjpi@iaa.es), N.L (Nikolai.Lehtinen@uib.no), A.L (aluque@iaa.es), or F.J.G.V (vazquez@iaa.es).

\end{document}